\documentclass[twocolumn,prb,aps,showpacs]{revtex4}
\usepackage[dvips]{graphicx}
\begin{document}
\title{Sum rules and vertex corrections for electron-phonon interactions}
\author{O.~R\"osch}\altaffiliation{Present address: Institut f\"ur Theoretische Physik, Universit\"at zu K\"oln, Z\"ulpicher Str.\ 77, D-50937 K\"oln, Germany}
\author{G.~Sangiovanni}
\author{O.~Gunnarsson}
\affiliation{
Max-Planck-Institut f\"ur Festk\"orperforschung, Heisenbergstr.\ 1, D-70506 Stuttgart, Germany
}

\begin{abstract}
We derive sum rules for the phonon self-energy and the electron-phonon
contribution to the electron self-energy of the Holstein-Hubbard model in the
limit of large Coulomb interaction $U$. Their relevance for finite $U$ is investigated using exact diagonalization and dynamical mean-field theory. Based on these sum rules, we study the importance of vertex corrections to the electron-phonon interaction in a diagrammatic approach. We show that they are crucial for a sum rule for the electron self-energy in the undoped system while a sum rule related to the phonon self-energy of doped systems is satisfied even if vertex corrections are neglected. We provide explicit results for the vertex function of a
two-site model.
\end{abstract}


\maketitle

\section{Introduction}
Recently, there has been much interest in the possibility 
that electron-phonon interactions may play an important 
role for properties of cuprates, e.g., for 
superconductivity.\cite{Shen,Pint,Gweon} In particular, the 
interest has focused on the idea that the Coulomb interaction
$U$ might enhance effects of electron-phonon interactions,
e.g., due to interactions with spin fluctuations.\cite{Mishchenko}
Effects of the electron-phonon coupling are described by the electron-phonon part 
$\Sigma^{\rm ep}$ of the electron self-energy $\Sigma$ and 
the phonon self-energy, $\Pi$. We have previously derived sum 
rules for these quantities for the $t$-$J$ model, and used the 
sum rules to demonstrate that the electron-phonon interaction influences $\Sigma^{\rm ep}$ 
and $\Pi$ in quite different ways for strongly correlated 
systems.\cite{Oliversum} Here, we extend this work and derive sum 
rules for the related half-filled Holstein-Hubbard model in the limit of 
a large $U$. We obtain sum rules for $\Sigma^{\rm ep}$ 
integrating either over all frequencies or only over frequencies 
in the photoemission energy range. The latter sum rule 
shows a very strong dependence on $U$, increasing by a factor of 
four in going from $U=0$ to $U=\infty$. From numerical calculations,
using both exact diagonalization and dynamical mean-field theory,
we show that the $U=\infty$ result is also relevant for intermediate 
values of $U\approx 3D$, where $D$ is half the band width.  

In a diagrammatic many-body language, the electron-phonon interaction could be enhanced
by $U$ via corrections to vertex functions \cite{Huang,Koch} or dressing
of Green's functions.\cite{Ramsak} Huang {\it et al.} \cite{Huang} 
and Koch and Zeyher \cite{Koch} studied how $U$ changes an effective 
vertex function in the static limit ($\omega=0$) and found a 
suppression, although it was concluded in Ref.~\onlinecite{Huang}     
that the suppression is reduced for a large $U$ and a small 
wave vector ${\bf q}$. Often, one is not only interested in these 
special cases but in properties that depend on integrals over $\omega$ 
and ${\bf q}$ containing vertex functions. Here, we study to what extent
the sum rules above are fulfilled when vertex corrections are neglected.  
We find that the sum rule for $\Sigma^{\rm ep}$, integrating over the 
photoemission energy range, is underestimated by a factor of four if 
vertex corrections are neglected. On the other hand, a sum 
rule for the phonon self-energy is fulfilled also without 
vertex corrections. This suggests that it can be important to 
include vertex corrections for studying properties of cuprates 
and other strongly correlated materials.

The Hubbard model with electron-phonon interaction is introduced in Sec.~\ref{sec_model}. In Sec.~\ref{sec_sr}, sum rules for the electron and phonon self-energies are derived focusing on the limit $U\to\infty$ and in Sec.~\ref{sec:num} we numerically check their accuracy for large but finite $U$. These sum rules then form the basis for the discussion of the effects of vertex corrections in Sec.~\ref{sec_eff}. The results are illustrated in Sec.~\ref{sec_ex} for a two-site model.

\section{Hubbard model in the limit of large $U$}\label{sec_model}
Strongly correlated electrons are often described by the Hubbard model
\begin{equation}\label{eq:1}
H=\varepsilon_d\sum_{i,\sigma}n_{i\sigma}-
t\sum_{\langle i,j\rangle,\sigma}(c^{\dagger}_{i\sigma}
c^{\phantom \dagger}_{j\sigma}+{\rm H.c.})
+U\sum_{i}n_{i\uparrow}n_{i\downarrow},
\end{equation}
where $\varepsilon_d$ is the level energy, $t(>0)$ is the hopping
integral between nearest-neighbor sites $\langle i,j\rangle$, $U$ is the Coulomb repulsion between two electrons
on the same site, $c_{i\sigma}^{\dagger}$ creates an electron 
on site $i$ with spin $\sigma$, and $n_{i\sigma}=c_{i\sigma}^
{\dagger}c^{\phantom\dagger}_{i\sigma}$.
In addition, we introduce an electron-phonon interaction
\begin{equation}\label{eq:3}
H_{\rm ep}={1\over \sqrt{N}}\sum_{i,{\bf q}} g_{\bf q}(n_i-1)(b_{\bf q}^{\phantom\dagger}+
b_{-\bf q}^{\dagger})e^{i{\bf q}\cdot {\bf R}_i},
\end{equation}
where $N$ is the number of sites, $n_i=n_{i\uparrow}+
n_{i\downarrow}$, and $b^\dagger_{\bf q}$ creates a phonon with the wave vector ${\bf q}$ and energy $\omega_{\bf q}$ as described by the free phonon Hamiltonian
\begin{equation}\label{eq:Hph}
H_{\rm ph}=
\sum_{\bf q}
\omega_{\bf q}
b_{\bf q}^\dagger b_{\bf q}^{\phantom\dagger}.
\end{equation}
We assume a ${\bf q}$-dependent on-site 
coupling with the strength $g_{\bf q}$. The coupling to 
hopping integrals is neglected, which, e.g., has been found to be a good approximation for the planar oxygen (half-)breathing mode in the high-$T_\mathrm{c}$ cuprates.\cite{OlivertJ} The special case of a Holstein coupling is obtained by setting $g_{\bf q}=g$ and $\omega_{\bf q}=\omega_\mathrm{ph}$ for all ${\bf q}$.

To describe photoemission (PES) and inverse photoemission (IPES) within the sudden approximation, we consider the one-electron removal (-) and addition (+) spectra
\begin{eqnarray}
A^-({\bf k},\omega)=
\sum_{mn}\frac{e^{-\beta\mathcal{E}_m}}{Z}
|\langle n|c^{\phantom\dagger}_{{\bf k}\sigma}|m\rangle|^2
\delta(\omega\!+\!\mathcal{E}_n\!-\!\mathcal{E}_m),
\\
A^+({\bf k},\omega)=
\sum_{mn}\frac{e^{-\beta\mathcal{E}_m}}{Z}
|\langle n|c^\dagger_{{\bf k}\sigma}|m\rangle|^2
\delta(\omega\!+\!\mathcal{E}_m\!-\!\mathcal{E}_n),
\end{eqnarray}
where the energy $\omega$ is measured relative to the chemical potential $\mu$, $|m\rangle$ and $|n\rangle$ are eigenstates of the grand canonical Hamiltonian $\mathcal{H}=H-\mu\langle N\rangle$ with eigenenergies $\mathcal{E}_{m/n}$, $Z=\sum_m\exp(-\beta\mathcal{E}_m)$ is the corresponding partition sum, and $\beta=1/T$.
We assume that there is no explicit dependence on the electron spin $\sigma$.
From the total spectral density
\begin{equation}\label{eq:tsd}
A({\bf k},\omega)=A^-({\bf k},\omega)+A^+({\bf k},\omega),
\end{equation}
we obtain the one-electron Green's function
\begin{equation}\label{eq:spec_repr}
G({\bf k},z)=
\int_{-\infty}^\infty\!d\omega
\ \frac{A({\bf k},\omega)}{z-\omega}
\end{equation}
which depends on the electronic wave vector $\bf k$ and the complex energy $z$. It is related to the electron self-energy $\Sigma({\bf k},z)$ via the Dyson equation
\begin{equation}\label{eq:Dyson_full}
G({\bf k},z)=\frac{1}{z-\varepsilon_{\bf k}-\Sigma({\bf k},z)},
\end{equation}
where $\varepsilon_{\bf k}=\varepsilon_d-t\sum_{\langle i,j\rangle}
(e^{i{\bf k}({\bf R}_i-{\bf R}_j)}+\mathrm{H.c.})/N$ is the bare electronic dispersion.
We can split up $G({\bf k},z)$ into two parts,
\begin{equation}\label{eq:splitup}
G({\bf k},z)=G^+({\bf k},z)+G^-({\bf k},z),
\end{equation}
where the (I)PES Green's functions $G^\pm({\bf k},z)$ are defined by replacing the spectral density in Eq.~(\ref{eq:spec_repr}) by $A^\pm({\bf k},\omega)$.

We also define corresponding self-energies by
\begin{equation}\label{eq:Dyson_partial}
G^\pm({\bf k},z)=\frac{a_{\bf k}^\pm}{z-\varepsilon_{\bf k}^\pm-\Sigma^\pm({\bf k},z)},
\end{equation}
where $a_{\bf k}^\pm$ is the integrated weight and $\varepsilon_{\bf k}^\pm$ absorbs energy-independent contributions to the self-energy. We consider the half-filled system and choose $\varepsilon_d=-U/2$ to have explicit particle-hole symmetry. Then, as detailed in the appendix, $a_{\bf k}^\pm\to1/2$, $\varepsilon_{\bf k}^\pm\to\pm U/2+\varepsilon_{\bf k}$ for $U\to\infty$. In this limit, one finds from inserting Eqs.~(\ref{eq:Dyson_full}) and (\ref{eq:Dyson_partial}) into Eq.~(\ref{eq:splitup}) to leading order in $1/U$
\begin{equation}\label{eq:relate}
\mathrm{Im}\ \Sigma({\bf k},z\approx\pm U/2)
=2\ \mathrm{Im}\ \Sigma^\pm({\bf k},z\approx\pm U/2),
\end{equation}
relating the spectral densities of the different self-energies for energies $z\approx\pm U/2$.

In the limit of large $U$, states with double 
occupancy can be projected out. This leads to the $t$-$J$ model
\cite{Zhang} as an effective low-energy model for
Eq.~(\ref{eq:1}) if certain terms are assumed to be negligible,\cite{Auerbach} as 
will be done here. Since double occupancy is excluded in the $t$-$J$ model, its electron Green's function has no contribution from inverse photoemission for the undoped system, i.e., $G_{t\textrm{-}J}({\bf k},z)=G^-_{t\textrm{-}J}({\bf k},z)$. We assume that the photoemission 
spectra of the Hubbard and the $t$-$J$ model are identical (apart from a trivial energy shift $\approx U/2$), and thus
\begin{equation}\label{eq:relminus}
G^-({\bf k},z)=G_{t\textrm{-}J}({\bf k},z+U/2).
\end{equation}
The inverse photoemission in the half-filled Hubbard model can be related to the photoemission in the undoped $t$-$J$ model because of particle-hole symmetry,
\begin{equation}\label{eq:relplus}
G^+({\bf k},z)=G_{t\textrm{-}J}({\bf k}_\mathrm{AF}-{\bf k},U/2-z)
\end{equation}
for a two-dimensional square lattice where ${\bf k}_\mathrm{AF}=(\pi/a,\pi/a)$. 
Terms of order $t\ll U$ are neglected in the energy shifts $\pm U/2$ in Eqs.~(\ref{eq:relminus}) and (\ref{eq:relplus}).

The electron self-energy in the $t$-$J$ model is defined via $G_{t\textrm{-}J}({\bf k},z)=0.5/[\omega-\Sigma_{t\textrm{-}J}({\bf k},z)]$ and it follows from Eqs.~(\ref{eq:Dyson_partial}), (\ref{eq:relminus}), and (\ref{eq:relplus}) that
\begin{eqnarray}\label{eq:relsigminus}
\Sigma^-({\bf k},z)&=&\Sigma_{t\textrm{-}J}({\bf k},z+U/2),\\
\label{eq:relsigminus2}
\Sigma^+({\bf k},z)&=&\Sigma_{t\textrm{-}J}({\bf k}_\mathrm{AF}-{\bf k},U/2-z).
\end{eqnarray}

\section{Sum rules}\label{sec_sr}

In this section, we discuss sum rules for the electron and phonon self-energies. They will allow us to address the importance of vertex corrections to the electron-phonon interaction in a diagrammatic treatment of the Hubbard model.

\subsection{Electron self-energy}\label{ssec:ese}

In App.~\ref{der_sum}, we use spectral moments to derive the following sum rule for the electron self-energy $\Sigma
({\bf k},z)$ which gives the total weight of its spectral density integrated over all frequencies:
\begin{eqnarray}\label{eq:fsr}
&&\hskip-0.3cm
\frac{1}{\pi}\int_{-\infty}^\infty
\!d\omega\ \mathrm{Im}\ \Sigma_\sigma
({\bf k},\omega-i0^+)
=U^2\langle n^{\phantom\dagger}_{i-\sigma}
\rangle\left(1-\langle n_{i-\sigma}\rangle\right)\nonumber\\
&&\hskip-0.2cm+2U\left[
\sum_{\bf q}\frac{g_{\bf q}}{\sqrt{N}}
\langle(b^{\phantom\dagger}_{\bf q}+b_{-{\bf q}}^\dagger)\rho_{{\bf q}-\sigma}\rangle
-g_{{\bf q}={\bf 0}}\langle b^{\phantom\dagger}_i+b_i^\dagger\rangle
\langle n_{i-\sigma}\rangle
\right]\nonumber\\
&&\hskip-0.2cm+\frac{1}{N}
\sum_{\bf q}
|g_{\bf q}|^2\langle|b^{\phantom\dagger}_{\bf q}+b_{-{\bf q}}^\dagger|^2\rangle
-g_{{\bf q}={\bf 0}}^2\langle b^{\phantom\dagger}_i+b_i^\dagger\rangle^2,
\end{eqnarray}
where we defined $\rho_{{\bf q}\sigma}=\sum_in_{i\sigma}e^{-i{\bf q}{\bf R}_i}/\sqrt{N}$ and
$b_i=\sum_{\bf q}b_{\bf q}e^{i{\bf q}{\bf R}_i}/\sqrt{N}$.
$\sigma$ is the electron spin for which the self-energy is calculated but in our case the results do not depend on it.
% The index $\sigma$ for the electron spin is kept explicitly, but in most of the other sections we drop it because we assume spin-independent results \emph{How to formulate better?}.
The sum rule in Eq.~(\ref{eq:fsr}) is valid for any $U$ and interestingly, it is independent of the electronic wave vector $\bf k$.
For a Holstein coupling, it simplifies to
\begin{eqnarray}\label{eq:totalsumrule}
&&\hskip-0.5cm
\frac{1}{\pi}\int_{-\infty}^\infty
\!d\omega\ \mathrm{Im}\ \Sigma_\sigma
({\bf k},\omega-i0^+)
=U^2\langle n^{\phantom\dagger}_{i-\sigma}
\rangle\left(1-\langle n_{i-\sigma}\rangle\right)\nonumber\\
&&\quad\quad+2Ug\langle(b^{\phantom\dagger}_i+b_i^\dagger)n_{i-\sigma}\rangle
+g^2\langle(b^{\phantom\dagger}_i+b_i^\dagger)^2\rangle.
\end{eqnarray}
In the derivation of Eqs.~(\ref{eq:fsr}) and (\ref{eq:totalsumrule}), we have assumed translation invariance so the expectation values on the right hand side of the equations do not depend on the site $i$ at which they are evaluated.

We now focus on the half-filled system where the mean occupation per site and spin $\langle n_{i-\sigma}\rangle=1/2$ and consider the limit $U\to\infty$.
Because of the complete suppression of double occupancies and the specific form of the electron-phonon coupling in Eq.~(\ref{eq:3}), there are no phonons excited in the ground state and expectation values involving phonon operators in Eqs.~(\ref{eq:fsr}) and (\ref{eq:totalsumrule}) greatly simplify, e.g., $U\langle(b_i+b_i^\dagger)n_{i-\sigma}\rangle\to0$
and $\langle(b_i+b_i^\dagger)^2\rangle\to1$. For the electron-phonon contribution $\Sigma^\mathrm{ep}({\bf k},z)$ to the electron self-energy, i.e. the difference between the self-energies for systems with and without electron-phonon coupling, one then obtains the following sum rule:
\begin{equation}\label{eq:largeUsumrule}
\lim_{U\to\infty}\frac{1}{\pi}\int_{-\infty}^\infty
\!d\omega\ \mathrm{Im}\ \Sigma^\mathrm{ep}
({\bf k},\omega-i0^+)=\frac{1}{N}\sum_{\bf q}|g_{\bf q}|^2\equiv\overline g^2.
\end{equation}
In the special case of a Holstein coupling, $\overline g=g$.

As described in App.~(\ref{der_par}), corresponding sum rules for the electron-phonon contributions to the (I)PES self-energies $\Sigma^\pm({\bf k},z)$ can be derived analogously. One finds
\begin{equation}\label{eq:largeUsumrule2}
\lim_{U\to\infty}\frac{1}{\pi}\int_{-\infty}^\infty
\!d\omega\ \mathrm{Im}\ \Sigma^{\pm,\mathrm{ep}}
({\bf k},\omega-i0^+)=\overline g^2.
\end{equation}
It follows from Eqs.~(\ref{eq:relsigminus}) and (\ref{eq:relsigminus2}) that Eq.~(\ref{eq:largeUsumrule2}) directly translates into a sum rule for the electron-phonon contribution to the electron self-energy in the undoped $t$-$J$ model. This result was obtained already in Ref.~\onlinecite{Oliversum} using spectral moments in the framework of the $t$-$J$ model itself.

The spectral functions of $\Sigma^\pm({\bf k},z)$ are non-zero only in the energy range where the (I)PES spectra $\mathrm{Im}\ G^\pm
({\bf k},\omega-i0^+)/\pi$ are located, i.e., around $\omega\approx\pm U/2$.
With Eq.~(\ref{eq:relate}), we can therefore derive from Eq.~(\ref{eq:largeUsumrule2}) also partial sum rules for the electron-phonon contribution to the full self-energy $\Sigma({\bf k},z)$ in the Hubbard model,
\begin{equation}\label{eq:largeUsumrule3}
\lim_{U\to\infty}\frac{1}{\pi}\int_\mathrm{(I)PES}
\!d\omega\ \mathrm{Im}\ \Sigma^\mathrm{ep}
({\bf k},\omega-i0^+)=2\overline g^2,
\end{equation}
where we integrate over the (I)PES energy range around $\omega\approx\pm U/2$. 
An explicit choice in the limit of large $U$ would be, e.g., to integrate from $-\infty$ to $-U/4$ (from $U/4$ to $\infty$) in order to fully include the energy range of the PES (IPES) spectrum.

Together with Eq.~(\ref{eq:largeUsumrule}), it follows from Eq.~(\ref{eq:largeUsumrule3}) that the corresponding partial sum rule for integrating over the central energy range $\omega\approx0$ is given by
\begin{equation}\label{eq:largeUsumrule4}
\lim_{U\to\infty}\frac{1}{\pi}\int_{\omega\approx0}
\!d\omega\ \mathrm{Im}\ \Sigma^\mathrm{ep}
({\bf k},\omega-i0^+)=-3\overline g^2.
\end{equation}
The negative value can be understood as follows. For 
$\bar g=0$, $\mathrm{Im}\ \Sigma({\bf k},\omega\!-\!i0^+)/\pi$ has a pole in this energy range with a large positive weight ($\approx U^2/4$). When the electron-phonon coupling is switched on, the strength of this pole is slightly reduced which shows up as a pole with a negative weight in the spectral function of  $\Sigma^\mathrm{ep}({\bf k},z)$.

\begin{figure}
\centerline{
{\rotatebox{0}{\resizebox{8.5cm}{!}{\includegraphics {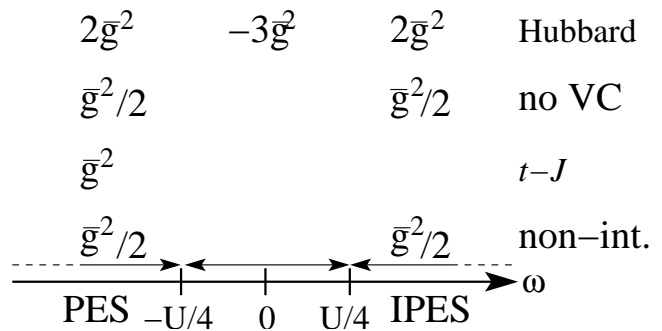}}}}}
\caption{\label{fig:1}
Weights obtained by integrating $\mathrm{Im}\ \Sigma^{\rm ep}({\bf k},
\omega-i0^+)/\pi$ over the indicated frequency intervals 
for the half-filled Hubbard and undoped $t$-$J$ models. Also 
shown are the result for the Hubbard model without vertex 
corrections (no VC) and the lowest-order result for the 
Hubbard model with $U=0$ (non-int.). For the $t$-$J$ 
model, the photoemission spectrum has been shifted by -$U/2$ 
and for the $U=0$ Hubbard model the photoemission and inverse 
photoemission spectra have been shifted by -$U/2$ and 
$U/2$, respectively. The results for $U=0$ refer to the
${\bf k}$-averaged self-energy.
}
\end{figure}

The different sum rules for the electron-phonon contribution to the self-energy in the Hubbard and the $t$-$J$ model are summarized schematically in Fig.~\ref{fig:1}.
For comparison, we also show the result for non-interacting electrons ($U=0$) to lowest order in $\overline g^2$.
Integrating the ${\bf k}$-averaged $\Sigma^{\rm ep}_{\mathrm{non}\textrm{-}\mathrm{int}}$ over 
the range of the photoemission spectrum for a half-filled 
model gives the contribution $\bar g^2/2$, which is a factor 
of 4 smaller than what is obtained in the large-$U$ model (cf.\ Eq.~(\ref{eq:largeUsumrule3})). Results for the large-$U$ Hubbard model obtained by neglecting vertex corrections (no VC) are also shown in Fig.~\ref{fig:1} and will be discussed in Sec.~\ref{sec_eff}.

\subsection{Phonon self-energy}

To lowest order in $g_{\bf q}$, the phonon self-energy $\Pi({\bf q},z)$ is given by $|g_{\bf q}|^2\chi({\bf q},z)/N$ where $\chi({\bf q},z)$ is the charge-charge response function. Sum rules for $\chi({\bf q},z)$ therefore directly translate into approximate sum rules for $\Pi({\bf q},z)$. At $T=0$,
\begin{equation}\label{eq:12a}
\chi({\bf q},z)=\sum_{\nu}|\langle \nu|\rho_{\bf q}
|0\rangle|^2\left({1\over z-\omega_{\nu}
}-{1\over z+\omega_{\nu}} \right),
\end{equation}
where $|\nu\rangle$ is an eigenstate of $H$ with eigenenergy $\omega_{\nu}$ relative to the ground state energy and $\rho_{\bf q}=\sum_\sigma\rho_{{\bf q}\sigma}$ is the    
Fourier transform of $n_i$.
For $U\gg |t|$ and a half-filled system, the ground state has
exactly one electron per site to lowest order in $t/U$. 
Applying $\rho_{\bf q}$ to the ground state $|0\rangle$, 
it then follows that $\rho_{\bf q}|0\rangle$ is zero
to this order if we consider ${\bf q}\neq{\bf 0}$. The sum rule for $|\mathrm{Im}\ \chi({\bf q},\omega+i0^+)|/(\pi N)$
integrated over all frequencies is then also zero to lowest 
order in $t/U$:
\begin{equation}\label{eq:chsr}
{1\over \pi N}\int_{-\infty}^{\infty}\!d\omega 
\ |\mathrm{Im}\ \chi( {\bf q}\neq{\bf 0},\omega+i0^+)|
=\mathcal{O}(t/U).
\end{equation}

In the $t$-$J$ model, an exact sum rule for the spectral function of $\chi({\bf q},z)$ has been derived:\cite{Horsch1}
\begin{equation}\label{eq:8}
{1\over \pi N^2}\sum_{{\bf q}\ne 0}\int_{-\infty}^{\infty}\!d\omega 
\ |\mathrm{Im}\ \chi_{t\textrm{-}J}( {\bf q},\omega+i0^+)|
=2\delta(1-\delta),
\end{equation}
where $\delta$ is the doping.
We assume as before that the photoemission spectra of the large-$U$ Hubbard and the $t$-$J$ model are identical. They extend over an energy range which is equal or smaller than the width of the lower Hubbard band, $\Delta=\mathcal{O}(t)\ll U$. Then, Eq.~(\ref{eq:8}) also leads to a sum rule for $\chi({\bf q},z)$ if the integration is limited to $|\omega|\le2\Delta$ which excludes transitions from the lower to the upper Hubbard bands not captured by the $t$-$J$ model:
\begin{equation}\label{eq:8a}
{1\over \pi N^2}\sum_{{\bf q}\ne 0}\int_{-2\Delta}^{2\Delta}\!d\omega 
\ |\mathrm{Im}\ \chi( {\bf q},\omega+i0^+)|
=2\delta(1-\delta).
\end{equation}
As discussed in Ref.~\onlinecite{Oliversum}, these results indicate a strong suppression of the phonon self-energy in weakly doped systems with strong correlations because one would obtain unity if non-interacting electrons ($U=0$) were assumed instead.

\section{Numerical results at finite $U$}\label{sec:num}

The sum rules in Sec.~\ref{ssec:ese} for the electron-phonon contribution to the electron self-energy have been derived in the limit $U\to\infty$. In practice, however, we are interested in strongly correlated systems with large but finite $U$. In order to check the usefulness of the sum rules for such cases we have performed numerical calculations using exact diagonalization (ED) and dynamical mean-field theory (DMFT).

The ED calculations are done on a two-dimensional tilted 10-site square cluster with periodic boundary conditions. We consider a weak electron-phonon coupling such that it suffices to include only states with at most one phonon excited, thereby limiting the size of the phonon Hilbert space. In addition to these exact results that are only influenced by finite size effects, we also consider the thermodynamic limit in the dynamical mean-field approximation\cite{Geo96} which in addition allows us to consider larger electron-phonon couplings. Instead of studying their effects in the often assumed paramagnetic phase,\cite{San05} we do DMFT in the antiferromagnetic phase (AF-DMFT) since antiferromagnetic correlations are believed to be important.\cite{San06} We consider a Bethe lattice with a semi-elliptical density of states (half bandwidth $D$). The self-consistent Anderson impurity model which appears in this approach is solved using exact diagonalization and a continued fraction expansion\cite{Si94} (employing up to 14 discrete bath levels and allowing for up to 30 excited phonons).
Specifically, we study the undoped Hubbard-Holstein model with $D=4t=1$ and $\omega_\mathrm{ph}=0.025D=0.1t$.
In both approaches, we first calculate the (I)PES spectral densities $A^\pm$ which directly lead to the Green's function $G$ using Eqs.~(\ref{eq:tsd}) and (\ref{eq:spec_repr}). The inversion of the Dyson equation (Eq.~(\ref{eq:Dyson_full})) then gives the self-energy $\Sigma$ whose spectral density we integrate over different energy ranges. Alternatively, the total sum rule can be obtained from calculating the ground state expectation values appearing in Eq.~(\ref{eq:totalsumrule}).

We first consider a weak electron-phonon coupling corresponding to a dimensionless coupling constant $\lambda=g^2/(\omega_\mathrm{ph}D)=g^2/(\omega_\mathrm{ph}4t)=0.0025$.
\begin{figure}
\includegraphics[width=8cm]{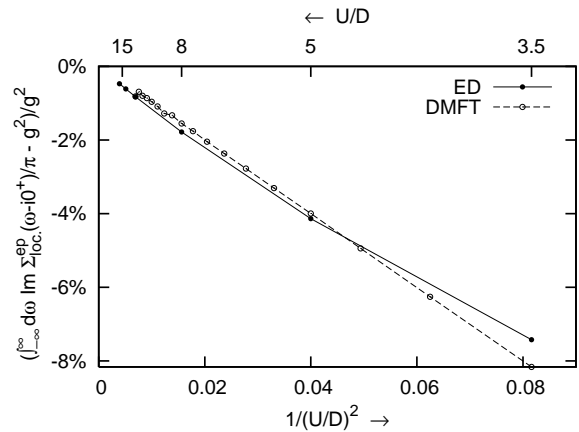}
\caption{\label{fig:comparison}
Relative deviation of the total sum rule for the local ($\bf k$-averaged) $\Sigma^\mathrm{ep}$ from its large-$U$ limit (Eq.~(\ref{eq:largeUsumrule})) as a function of $U$ using ED and DMFT with $\lambda=0.0025$.
}
\end{figure}
Using results from both ED and DMFT calculations, Fig.~\ref{fig:comparison} shows how much the total spectral weight of the electron-phonon contribution to the local ($\bf k$-averaged) electron self-energy at finite $U$ deviates from the sum rule for $U\to\infty$ (Eq.~(\ref{eq:largeUsumrule})). Relative to the latter, the deviation is less than 10\% for $U$ as small as $3.5D$, a value often considered appropriate for the cuprates. This difference decreases like $1/U^2$ as can be seen from Fig.~\ref{fig:comparison}, in agreement with expectations from a simple perturbational approach which we discuss in more detail at the end of this section. We note that results from ED and DMFT agree rather well. This consistency indicates that the finite size effects of the former and the approximations of the latter method are probably not strongly influencing the results discussed here.

\begin{figure}
\includegraphics[width=8cm]{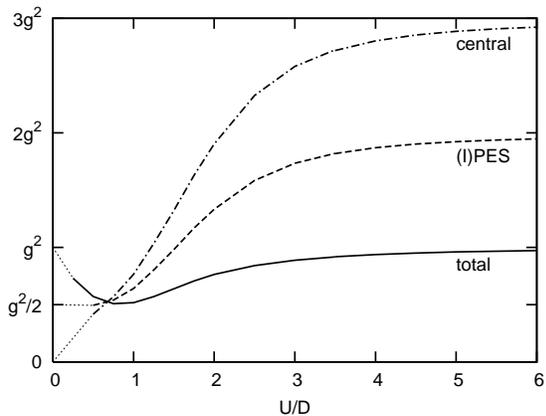}
\caption{\label{fig:smallU}
$U$ dependence of total and partial sum rules for $\Sigma^\mathrm{ep}$ using DMFT with $\lambda=0.0025$. The dotted lines indicate the expected small-$U$ behavior.
}
\end{figure}

The full $U$ dependence of both total and partial sum rules is illustrated in Fig.~\ref{fig:smallU} using results from DMFT calculations.
Already for $U$ larger than the non-interacting bandwidth $2D$, the sum rules clearly approach their respective large-$U$ limits (Eqs.~(\ref{eq:largeUsumrule}), (\ref{eq:largeUsumrule3}), and (\ref{eq:largeUsumrule4})).
As discussed at the end of Sec.~\ref{ssec:ese}, we expect for $U\to0$ that the total sum rule again approaches $g^2$ with the weight equally distributed over the PES and IPES energy ranges and no central contribution to the spectrum of $\Sigma^\mathrm{ep}$ at $\omega=0$. These trends are manifest in the calculated results.

Next, we study the dependence on the strength of the electron-phonon coupling for fixed $U=5D$. We restrict ourselves to DMFT calculations where larger couplings are accessible because of the smaller phonon Hilbert space compared to ED.
\begin{figure}
\includegraphics[width=8cm]{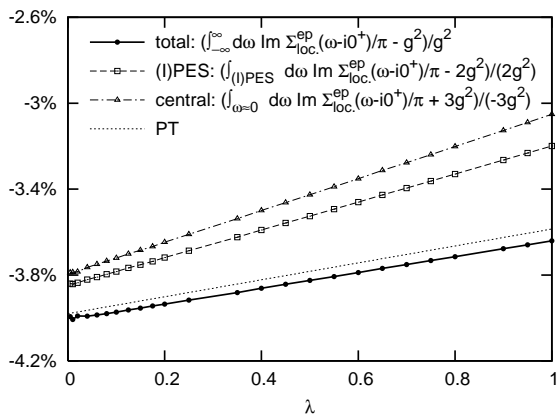}
\caption{\label{fig:lambda_dep}
Relative deviations of total and partial sum rules from their large-$U$ limits as a function of $\lambda$ using DMFT for $U=5D$. For the total sum rule, also results from perturbation theory (PT) for a simplified model (see text) are shown.
}
\end{figure}
In Fig.~\ref{fig:lambda_dep}, we plot the $\lambda$ dependence of the relative deviations of total and partially integrated spectral weights of the electron-phonon contribution to the local electron self-energy from their respective sum rules for $U\to\infty$ (Eqs.~(\ref{eq:largeUsumrule}), (\ref{eq:largeUsumrule3}), and (\ref{eq:largeUsumrule4})). For both the (I)PES and the central sum rules, the relative deviations are comparable in size to that of the total sum rule (this similarity is also found when the $U$ dependence is considered) and are less than 10\%. In all cases, the deviations decrease linearly with $\lambda$. Again, for the total sum rule, the results can be quite well described by a result from perturbation theory for a simplified model which we introduce in the following.

We consider the self-consistent Anderson impurity model to be solved in AF-DMFT of the Hubbard-Holstein model and replace the bath by a single (but spin-dependent) level. With the impurity level at $-U/2$, the self-consistent bath level is located at $\approx\pm U/2$ depending on the spin orientation; the hopping amplitude between impurity and bath is fixed to $V\approx D/2$. We treat both this hybridization and the electron-phonon interaction as perturbations of the atomic limit and find for large $U$
\begin{equation}\label{eq:pt1}
\langle(b^{\phantom\dagger}_i+b_i^\dagger)n_{i-\sigma}\rangle=-\frac{2V^2g}{U^2(U+\omega_\mathrm{ph})}+\mathcal{O}(U^{-4})
\end{equation}
and
\begin{equation}\label{eq:pt2}
\langle(b^{\phantom\dagger}_i+b_i^\dagger)^2\rangle=1+\frac{4V^2g^2/\omega_\mathrm{ph}}{U(U+\omega_\mathrm{ph})(U+2\omega_\mathrm{ph})}
+\mathcal{O}(U^{-4}).
\end{equation}
Using these results in Eq.~(\ref{eq:totalsumrule}), we expect the total spectral weight of the electron-phonon contribution to the electron self-energy to be proportional to $U^{-2}$ as was observed in Fig.~\ref{fig:comparison}. Although we have replaced the bath by a single level, the expressions in Eqs.~(\ref{eq:pt1}) and (\ref{eq:pt2}) give a rather accurate description of the numerical results as can be seen in Fig.~(\ref{fig:lambda_dep}).

In conclusion, we find that for typical values of $U$ the relative deviations from the sum rules derived for $U\to\infty$ are smaller than 10\%, and therefore these sum rules can be used semi-quantitatively also for finite $U$.

\section{Effects of vertex corrections}\label{sec_eff}

\begin{figure}
\centerline{
{\rotatebox{0}{\resizebox{4.0cm}{!}{\includegraphics {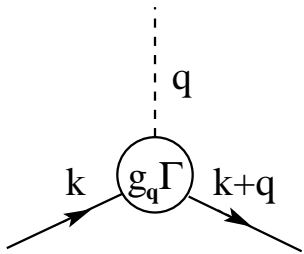}}}}}
\caption{\label{fig:0}
Diagrammatic representation of the vertex function $\Gamma(k,q)$, where $k$ ($q$) stands for the incoming electron (phonon) momentum and frequency. The full and 
dashed lines represent electron and phonon Green's 
functions, respectively. 
}
\end{figure}

We now use the sum rules introduced in Sec.~\ref{sec_sr} to study the effects of vertex corrections in a diagrammatic calculation of the electron and phonon self-energies. We define the vertex function $\Gamma(k,q)$ as the sum of all irreducible vertex diagrams connecting two electron Green's functions with a phonon propagator taking out one coupling constant $g_{\bf q}$ explicitly (see Fig.~\ref{fig:0}). We use the 4-vectors $k$ and $q$ as shorthand notation for the momenta and frequencies involved.

\subsection{Electron self-energy}
\begin{figure}
\begin{minipage}{5.5cm}
{\rotatebox{0}{\resizebox{6.0cm}{!}{\includegraphics {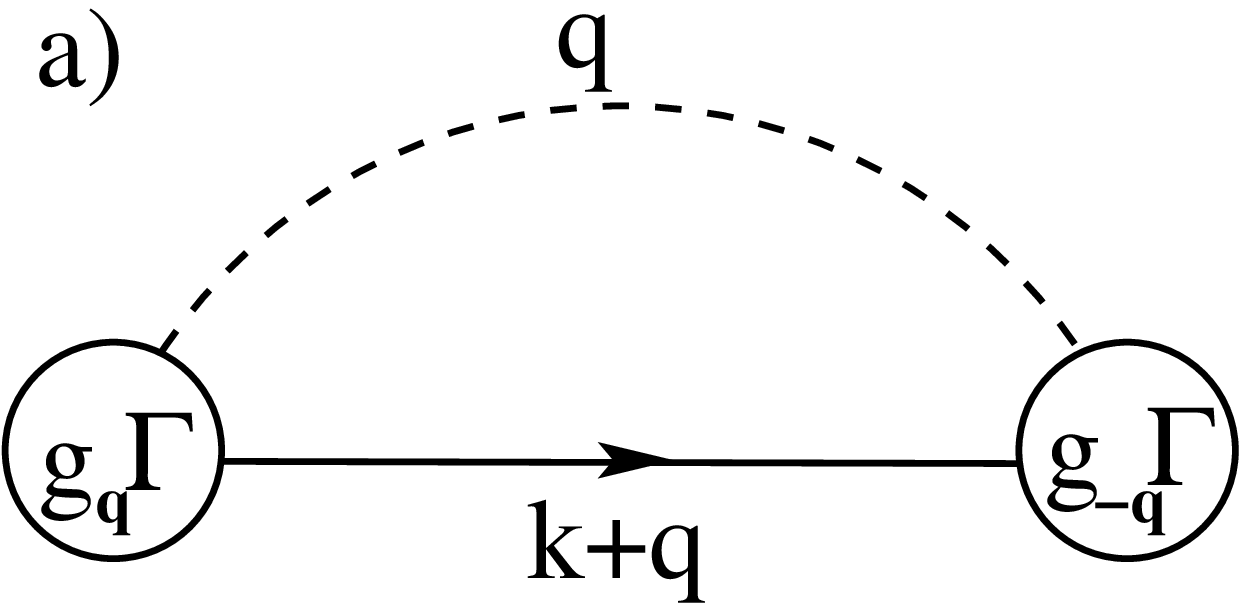}}}}
\end{minipage}

\begin{minipage}{5.5cm}
\vskip0.6cm
{\rotatebox{0}{\resizebox{5.2cm}{!}{\includegraphics {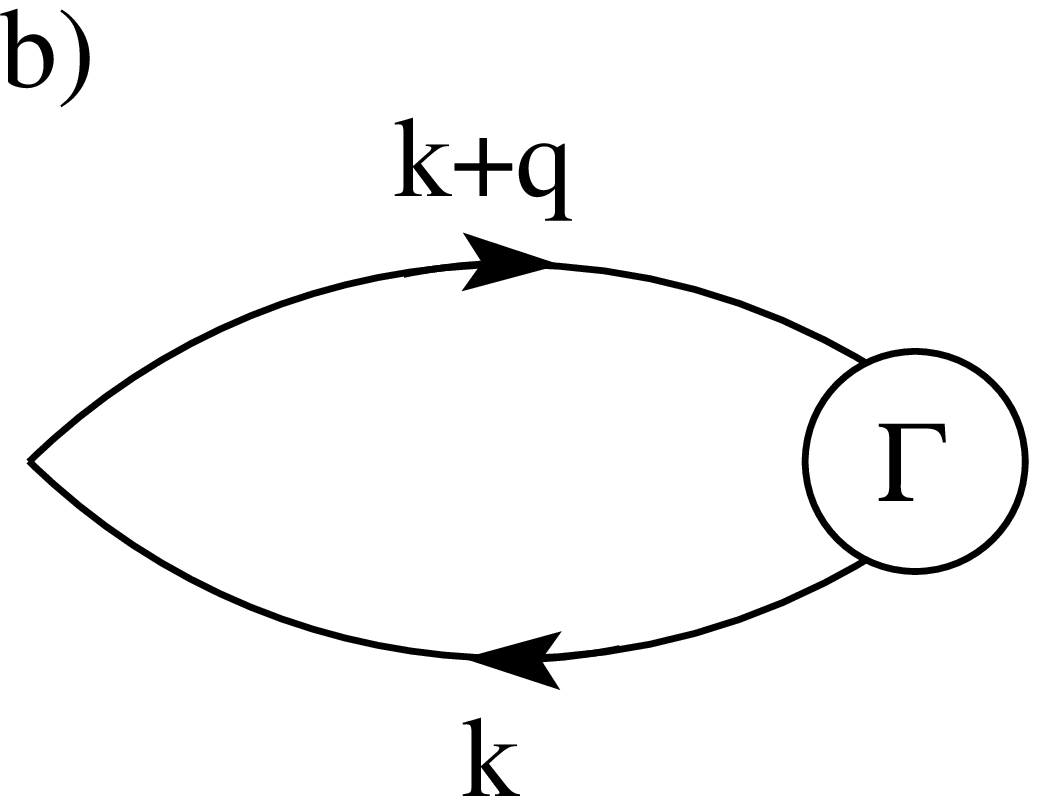}}}}
\end{minipage}
\caption{\label{fig:2}
a) Lowest-order contribution (in 
$|g_{\bf q}|^2$) to the electron-phonon part of 
the electron self-energy $\Sigma^\mathrm{ep}$ and b) the charge-charge 
response function $\chi$. The full and dashed lines represent 
electron and phonon Green's functions, respectively, and the 
circles the vertex functions $(g_{\bf q})\Gamma$. 
}
\end{figure}

An important lowest-order (in 
$|g_{\bf q}|^2$) contribution to the electron-phonon part of the electron self-energy is
shown in Fig. \ref{fig:2}a, although there are also other, more complicated lowest-order diagrams.\cite{Born} The 
diagram in Fig. \ref{fig:2}a is
\begin{eqnarray}\label{eq:9}
&&\!\!\!\!\!\Sigma^{\rm ep}({\bf k},\omega)= \\
&&{i\over N}\sum_{\bf q}|g_{\bf q}|^2
\int {d\omega'\over 2 \pi}\ G(k+q)D(q) \Gamma(k,q)\Gamma(k+q,-q), \nonumber
\end{eqnarray}
where $q$ stands for a wave vector ${\bf q}$ and a frequency
 $\omega'$. $G$ and $D$ are fully dressed electron and phonon Green's functions.

We now  neglect the vertex corrections, i.e, we put $\Gamma=1$. 
After using Eq.~(\ref{eq:spec_repr}) to express the electron Green's function in terms of its spectral 
function, the $\omega'$ integral can be performed. For the 
half-filled Hubbard model in the large-$U$ limit, the spectral 
function integrated over the lower or the upper Hubbard band gives half an electron 
per spin. As a result, we find      
\begin{equation}\label{eq:10}
\lim_{U\to\infty}
{1\over \pi} \int_\mathrm{(I)PES}\!d\omega\ \mathrm{Im}\ \Sigma^{\rm ep}_
{\rm no\ VC}({\bf k}, \omega-i0^{+})= {1\over 2}\bar g^2,
\end{equation}
where we have also used that at half-filling, the phonon Green's function $D(q)$ is not dressed in the
large-$U$ limit. Comparing this result for the diagram in Fig.~\ref{fig:2}a
without vertex corrections (no VC) with the corresponding exact sum rule, Eq.~(\ref{eq:largeUsumrule3}), shows that this approximation underestimates the sum rule by a factor of four. This result is schematically 
indicated in Fig.~\ref{fig:1}. When vertex corrections are neglected, the diagram in Fig.~\ref{fig:2} has no contributions in the energy range $\omega\approx0$.

We have elsewhere\cite{Born} used the self-consistent Born approximation to study electron-phonon interaction in the undoped $t$-$J$ model which is closely related to the half-filled Hubbard model in the large-$U$ limit. In this approach, fairly good agreement with exact results is obtained although vertex corrections are neglected. Already the lowest-order (in the electron-phonon coupling) diagram for the electron-phonon contribution to the electron self-energy fulfills an exact sum rule for the total spectral weight. The reason for this result contrasting the strong violation of the sum rule in the large-$U$ Hubbard model when vertex corrections are neglected can be traced to the use of a Green's function for spinless holons in the self-consistent Born approximation. Its spectral function integrates to unity over the photoemission energy range whereas the spin-dependent electron Green's function in the large-$U$ Hubbard model has the spectral weight one half in both the photoemission and the inverse photoemission energy range.

\subsection{Phonon self-energy}
For ${\bf q}\ne{\bf 0}$, the charge-charge response function,
$\chi({\bf q},\omega)$, for the Hubbard model can be 
obtained from the diagram in Fig. \ref{fig:2}b:
\begin{equation}\label{eq:11}
\chi({\bf q},\omega)=-2i\sum_{\bf k}\int 
{d\omega'\over 2\pi}\ G(k+q)G(k)\Gamma(k+q,-q).
\end{equation}
We consider the large-$U$ limit for a hole-doped Hubbard 
model. The ${\bf k}$-averaged photoemission spectrum for a 
given spin integrates to $(1-\delta)/2$. As earlier, we               
use the assumption that this spectrum agrees with the 
spectrum of the $t$-$J$ model and that it only extends over an energy range $\Delta\ll U$. In inverse 
photoemission, the probability of adding an electron to 
an unoccupied site is $\delta$. We therefore assume that $U$ 
is so large that the integral of the inverse photoemission 
spectrum for a given spin up to $\Delta$  
is given by $\delta$. We neglect vertex corrections 
and replace $\Gamma(p+q,-q)$ by unity. Introducing 
spectral functions, the $\omega'$ integration is performed. 
We consider the integral of the spectrum of the charge-charge response function over  
$|\omega|\le 2\Delta$, thereby excluding transitions between the two Hubbard bands. Then,
\begin{eqnarray}\label{eq:12}
&&\!\!\!\!\!{1\over \pi N^2}\sum_{{\bf q}}\int_{-2\Delta}^{2\Delta}
\!d\omega\ |{\rm Im}\ \chi_{\rm no\ VC}({\bf q},\omega+i0^+)|\nonumber \\
&&={2\over N^2}\sum_{\bf q}\sum_{\bf k}\lbrack w_{\rm P}({\bf k})
w_{\rm IP}({\bf k+q})+ w_{\rm P}({\bf k}+{\bf q})w_{\rm IP}({\bf k})\rbrack
\nonumber  \\ &&= 2\delta(1-\delta),
\end{eqnarray}
where $w_{\rm IP}({\bf k})$ is the integrated weight of the 
photoemission spectrum for the wave vector ${\bf k}$ and
$w_{\rm IP}({\bf k})$ is the corresponding quantity for inverse 
photoemission, excluding the upper Hubbard band. For 
a large system, the ${\bf q}={\bf 0}$ term gives a negligible 
contribution in Eq.~(\ref{eq:12}). It then agrees with the 
sum rule in Eq.~(\ref{eq:8a}), derived from the corresponding sum rule in the $t$-$J$ model, 
although vertex corrections have been neglected. 
It is important, however, to use dressed
Green's functions in calculating $\chi({\bf q},\omega)$.
Otherwise, $2\delta$ in Eq.~(\ref{eq:12}) would have been
replaced by $(1+\delta)$, and there would have been a strong 
disagreement with Eq.~(\ref{eq:8a}) in the low-doping regime.

The sum rule in Eq.~(\ref{eq:12}) refers to an average 
over ${\bf q}$. We next study individual values of ${\bf q}$ for a
$t$-$J$ model on a two-dimensional tilted 18-site square cluster
with two holes, periodic boundary conditions, and $J/t=0.3$. We have calculated $w_{\rm P}({\bf k})$ and $w_{\rm IP}({\bf k})$, using exact diagonalization. From
the second line of Eq.~(\ref{eq:12}), we can obtain sum 
rules for each  ${\bf q}$ similar to Eq.~(\ref{eq:12}). 
The result is shown in the line ``No VC'' of Table 
\ref{table:1}.  The results are compared with the sum 
rule for $\chi_{t\textrm{-}J}({\bf q},z)$ 
calculated in the $t$-$J$ model (``Exact'' in Table 
\ref{table:1}). As can be seen from the ratios of these results, the sum rule for individual values of ${\bf q}$ that can be deduced from Eq.~(\ref{eq:12}) is also rather well fulfilled 
(typically, with a deviation of 5-10\%) although vertex corrections 
are neglected. 

\begin{table}
\caption{\label{table:1}Sum rules equivalent to Eq.~(\ref{eq:12}) but for individual values of ${\bf q}$, using ED of the $t$-$J$ model on an 18-site cluster with two holes and $J/t=0.3$.
The ``No VC'' results were obtained from $w_{\rm P}({\bf k})$ and 
$w_{\rm IP}({\bf k})$, using the second line of 
Eq.~(\ref{eq:12}), and the exact ones result from the direct calculation of $\chi$. ``Ratio'' shows the ratio of these results. 
}
\begin{tabular}{lllllll}
\hline
\hline
%${\bf q}/{\pi\over 3}$ & $(0,0)$ & $(1,1)$ & $(2,0)$ & $(2,2)$ & 
${\bf q}/{\pi\over 3}$ & $(1,1)$ & $(2,0)$ & $(2,2)$ & 
$(3,1)$ & $(3,3)$ \\
\hline
%No VC & 0.1660 & 0.1848 & 0.1927 & 0.2103 & 0.2025 & 0.2285 \\
%Exact   & 28.44  & 0.2100 & 0.1961 & 0.2191 & 0.2085 & 0.2212 \\
%Ratio &          & 0.8804 & 0.9825 & 0.9597 & 0.9714 & 1.0330 \\
No VC & 0.1848 & 0.1927 & 0.2103 & 0.2025 & 0.2285 \\
Exact & 0.2100 & 0.1961 & 0.2191 & 0.2085 & 0.2212 \\
Ratio & 0.8804 & 0.9825 & 0.9597 & 0.9714 & 1.0330 \\

\hline
\hline
\end{tabular}
\end{table}

For the half-filled Hubbard model, there is no contribution 
to Im $\chi({\bf q},\omega+i0^+)$ for $|\omega|\ll U$. We therefore 
focus on contributions for $|\omega|\approx U$. If $|{\rm Im} 
\chi_{\rm no\ VC}({\bf q},\omega+i0^+)|/(\pi N)$ is integrated   
over {\it all} frequencies, we obtain unity. The exact result in Eq.~(\ref{eq:chsr}), however, is zero to lowest order in $t/U$.
This dramatic disagreement shows the 
importance of vertex corrections in this case. 

\section{Example: Two-site model}\label{sec_ex}
To study the vertex corrections more explicitly, we consider 
a two-site Hubbard model. The electron-phonon interaction in 
Eq.~(\ref{eq:3}) can then be split in $q=0$ and $q=\pi$ terms. 
The $q=0$ term has just the rather trivial but important effect 
of convoluting the spectra with phonon satellites, while the 
$q=\pi$ term introduces dynamics, scattering electrons between 
bonding and antibonding orbitals. We therefore only keep the 
the more interesting $q=\pi$ term here.

Following Huang {\it et al.},\cite{Huang} we can calculate 
the vertex function explicitly for the two-site model in the limit 
of $U/t$ very large. We consider an incoming electron in the bonding
orbital (+) with the frequency $\omega$ scattered by the antibonding
($q=\pi$) phonon with the frequency $\omega'$ into the antibonding 
orbital (-) with the frequency $\omega+\omega'$ and obtain
\begin{eqnarray}\label{eq:13}
&&\Gamma(\omega,+;\omega')=\Gamma(\omega+\omega',-;-\omega') \\
&&={\omega(\omega+\omega')+\omega't+U^2/4
\over (\omega+t)(\omega+\omega'-t)}, \nonumber
\end{eqnarray} 
where various terms of higher order in $t/U$ have been neglected.

Using this result for the vertex function, we can calculate the diagram in Fig.~\ref{fig:2}a according to Eq.~(\ref{eq:9}). In the limit of large $U$, we find poles with weight $2g^2$ both at $\omega\approx -U/2$ and at $\omega\approx U/2$. Therefore, the sum rules for integrating over either the PES or the IPES energy range, Eq.~(\ref{eq:largeUsumrule3}), are exactly fulfilled. Without vertex corrections, these sum rules are underestimated by a factor of four, cf.\ Eq.~(\ref{eq:10}). This can be understood by noting that for electronic energies in the range of the lower or upper Hubbard band, $|\omega|\approx U/2$, it follows from Eq.~(\ref{eq:13}) that 
\begin{equation}\label{eq:14}
\Gamma(|\omega|\approx{U\over 2},+;\omega')\approx 2
\end{equation}
when the phonon frequency $\omega'$ is assumed to be small compared to $U$. Therefore, including vertex corrections effectively increases the weight of poles around $|\omega|\approx U/2$ by a factor $\Gamma^2=4$. In addition, the self-energy calculated using vertex corrections also has poles at $\omega=t-\omega_\mathrm{ph}$ and $\omega=-t$, the latter being a double pole. Except for a different sign, to leading order in $U$ they give the same contribution, $\mp g^2(U/2)^2/(2t-\omega_\mathrm{ph})^2$, to the integral over the spectral function of the self-energy. The sum of the two contributions, however, is not zero, but one finds, taking into account also terms which involve lower powers of $U$, that it equals $-3g^2$ as expected from Eq.~(\ref{eq:largeUsumrule4}). Therefore, also the sum rule over all frequencies in Eq.~(\ref{eq:largeUsumrule}) is fulfilled.

The sum rule for the charge-charge response function in Eq.~(\ref{eq:8a}) applies to finite dopings.
For the two-site model, this implies the uninteresting
filling one, for which there is no Coulomb interaction.
It is interesting, however, to study the sum rule over all 
frequencies for the half-filled two-site model. As expected, 
we find that the neglect of vertex corrections, incorrectly, 
gives contributions at $\omega\approx\pm U$ with weight $1/2$, respectively. Only when the vertex function from Eq.~(\ref{eq:13}) is used in Eq.~(\ref{eq:11}), $\chi(q=\pi,\omega)$ vanishes to lowest order in $t/U$. This is the correct result from Eq.~(\ref{eq:chsr}).
 
\section{Summary}
We have derived exact sum rules for the electron-phonon contribution $\Sigma^{\rm ep}$ to the
electron self-energy of the half-filled Holstein-Hubbard model in the limit of large $U$. In particular, we consider integrations both over 
all frequencies and over frequencies in the photoemission energy range.
Comparing results for $U=\infty$ and $U=0$, we find identical sum rules
when integrating over all frequencies but a difference by a factor of 
four when considering frequencies corresponding to photoemission only. 
Using different numerical methods, we find that these sum rules
are relevant also for systems with intermediate values of $U\approx3D$ that are 
typically of interest. These sum rules should be 
useful for testing approximate calculational schemes. 

We have also used sum rules for studying the importance of vertex 
corrections in a diagrammatic approach to properties in the 
Hubbard-Holstein model. For a weakly doped Hubbard-Holstein model 
in the large-$U$ limit, the phonon self-energy $\Pi$ is strongly 
reduced compared to the non-interacting case. This is described by  
a sum rule which integrates Im $\Pi$ over a finite frequency range such
that transitions between the Hubbard bands are not involved. This sum rule is 
satisfied if properly dressed Green's functions are used to calculate 
the phonon self-energy, even if vertex corrections are neglected. 
The energy dependence of Im $\Pi({\bf q},\omega)$ could, nevertheless, 
be substantially influenced by vertex corrections. For the half-filled 
system, we have to integrate Im $\Pi$ over all frequencies to obtain
a nontrivial sum rule. This sum rule is only satisfied if vertex 
corrections are included. 

The sum rule for Im $\Sigma^{\rm ep}$, integrating over frequencies 
corresponding to photoemission only, is violated by a factor of four if 
vertex corrections are neglected. These results have been illustrated by 
an explicit calculation of the vertex function in a two-site model.
We have studied integrated quantities where all values of $|{\bf q}|$ and 
$\omega$ enter, both in terms of their relative ratio and their absolute 
magnitude. Therefore, our findings cannot be directly compared with previous 
ones which focused on the static limit\cite{Huang,Koch} or on small $|{\bf q}|$ and $\omega$.\cite{Castellani} Our results show that the inclusion of 
vertex corrections can be essential to correctly describe effects of 
electron-phonon interaction in strongly correlated systems.

\acknowledgments
We acknowledge useful discussions with Erik Koch, Claudio Castellani, and Massimo Capone.

\appendix

\section{Derivation of sum rules using spectral moments}

\subsection{Total sum rule}\label{der_sum}

To derive the sum rule for the total integrated weight of the spectral density of the electron self-energy, Eq.~(\ref{eq:fsr}), we
expand $(z-\omega)^{-1}$ in Eq.~(\ref{eq:spec_repr}) in powers of
$1/z$. One obtains
\begin{equation}\label{eq:Gmom}
G({\bf k},z)=\sum_{m=0}^\infty
\frac{M^{(m)}_{{\bf k}}}{z^{m+1}},
\end{equation}
where the moments of the spectral density are defined as
\begin{equation}
M^{(m)}_{{\bf k}}=\int_{-\infty}^\infty
\!d\omega\ \omega^mA({\bf k}\omega).
\end{equation}
On the other hand, using the Heisenberg equations of motion for the time-dependent operators in the definition of the spectral density, these moments can also be obtained from \cite{Pot97}
\begin{equation}\label{eq:moments}
M^{(m)}_{{\bf k}}=
\langle\{\mathcal{L}^mc^{\phantom\dagger}_{{\bf k}\sigma},c^\dagger_{{\bf k}\sigma}\}\rangle,
\end{equation}
with $\mathcal{LO}=[\mathcal{O},H]$. $[\mathcal{O},\mathcal{O}']$ and $\{\mathcal{O},\mathcal{O}'\}$ denote the commutator and the anticommutator of two operators $\mathcal{O},\mathcal{O}'$, respectively.

For the Hubbard model with electron-phonon interaction which was defined in Eqs. (\ref{eq:1})-(\ref{eq:Hph}), one obtains from Eq.~(\ref{eq:moments})
\begin{equation}\label{eq:M0}
M^{(0)}_{{\bf k}}=1,
\end{equation}
\begin{equation}
M^{(1)}_{{\bf k}}=
\varepsilon_{\bf k}
+U\langle n_{i-\sigma}\rangle
+g_{{\bf q}={\bf 0}}\langle b^{\phantom\dagger}_i+b_i^\dagger\rangle
,
\end{equation}
and
\begin{eqnarray}\label{eq:M2}
M^{(2)}_{{\bf k}}
&=&\varepsilon_{\bf k}^2+U^2\langle n_{i-\sigma}\rangle
+\frac{1}{N}
\sum_{\bf q}
|g_{\bf q}|^2\langle|b^{\phantom\dagger}_{\bf q}+b_{-{\bf q}}^\dagger|^2\rangle\nonumber\\
&&
+2\varepsilon_{\bf k}U\langle n_{i-\sigma}\rangle
+2\varepsilon_{\bf k}g_{{\bf q}={\bf 0}}\langle b^{\phantom\dagger}_i+b_i^\dagger\rangle\nonumber\\
&&+2U\frac{1}{\sqrt{N}}\sum_{\bf q}
g_{\bf q}\langle(b^{\phantom\dagger}_{\bf q}+b_{-{\bf q}}^\dagger)\rho_{{\bf q}-\sigma}\rangle,\nonumber
\end{eqnarray}
where $\varepsilon_{\bf k}$, $\rho_{{\bf q}\sigma}$, and
$b_i$ have already been defined after Eqs.~(\ref{eq:Dyson_full}) and (\ref{eq:fsr}).

When the $1/z$ expansion of the self-energy,
\begin{equation}
\Sigma({\bf k},z)=\sum_{m=0}^\infty
\frac{C^{(m)}_{{\bf k}}}{z^m},
\end{equation}
is inserted into Eq.~(\ref{eq:Dyson_full}), a comparison with Eq.~(\ref{eq:Gmom}) leads to
\begin{equation}\label{eq:C1}
C^{(1)}_{{\bf k}}=M^{(2)}_{{\bf k}}/M^{(0)}_{{\bf k}}-\left(M^{(1)}_{{\bf k}}/M^{(0)}_{{\bf k}}\right)^2.
\end{equation}
As $C^{(1)}_{{\bf k}}$ corresponds to the zeroth moment
of the spectral density of $\Sigma({\bf k},z)$, we arrive at the %desired
sum rule, Eq.~(\ref{eq:fsr}), when Eqs.~(\ref{eq:M0})-(\ref{eq:M2}) are used in Eq.~(\ref{eq:C1}).

\subsection{Sum rules for (I)PES self-energies}\label{der_par}

The derivation of sum rules for the (I)PES self-energies $\Sigma^\pm({\bf k},z)$ can be done in full analogy with the previous section. Only the moments of the (I)PES spectra $A^\pm({\bf k},\omega)$ are now obtained from 
\begin{eqnarray}\label{eq:eins}
M^{+,(m)}_{{\bf k}}=
\langle(\mathcal{L}^mc^{\phantom\dagger}_{{\bf k}\sigma})c^\dagger_{{\bf k}\sigma}\rangle,
\\
\label{eq:zwei}
M^{-,(m)}_{{\bf k}}=
\langle c^\dagger_{{\bf k}\sigma}\mathcal{L}^mc^{\phantom\dagger}_{{\bf k}\sigma}\rangle
\end{eqnarray}
instead of Eq.~(\ref{eq:moments}). Because Eqs.~(\ref{eq:eins}) and (\ref{eq:zwei}) do not contain an anticommutator, much more complicated results are obtained for these moments. We do not list them explicitly. It turns out, however, that they simplify considerably once the limit $U\to\infty$ is taken. Using the analog to Eq.~(\ref{eq:C1}) and focusing on the electron-phonon contribution to the self-energies, one then arrives at the result in Eq.~(\ref{eq:largeUsumrule2}).

\end{document}